\documentclass[%
 reprint,
superscriptaddress,
showpacs,preprintnumbers,
 amsmath,amssymb,
 aps,
prb,
]{revtex4-1}

\usepackage{comment,color}
\usepackage{braket}

\usepackage{graphicx}
\usepackage{dcolumn}
\usepackage{bm}

\begin{document}


\title{Generic phase diagram for Weyl superconductivity in mirror-symmetric superconductors
}

\author{Ryo Okugawa}
\affiliation{%
 Department of Physics, Tokyo Institute of Technology, 2-12-1 Ookayama, Meguro-ku, Tokyo 152-8551, Japan
}%
\author{Takehito Yokoyama}%
\affiliation{%
 Department of Physics, Tokyo Institute of Technology, 2-12-1 Ookayama, Meguro-ku, Tokyo 152-8551, Japan
}%

\date{\today}

\begin{abstract}
We study topological phase transitions in three-dimensional odd-parity or noncentrosymmetric superconductors with mirror symmetry when time-reversal symmetry is broken. 
We construct a generic phase diagram for Weyl superconductivity in the mirror-symmetric superconductors. 
It is shown that Weyl superconductivity generally emerges between the trivial and the topological crystalline superconductor phases.
We demonstrate how a trajectory of the Weyl nodes determines the change in mirror Chern numbers in the topological phase transition. 
We also discuss a relationship between particle-hole symmetry and the trajectory of the Weyl nodes which realizes the topological crystalline superconductor phase.
\end{abstract}


\maketitle

\textit{Introduction.}
Thanks to interplay of topology and crystal symmetries,
novel topological phases have been suggested theoretically.
Topological crystalline insulators \cite{Fu11, Hsieh12, Alexandradinata14, Fang15, Shiozaki15, Wang16N}
and topological semimetals \cite{Yang14N, Fang15b, Wieder16, Bzduvsek16, Bradlyn16, Watanabe16} are understandable as a manifestation of their complex interplay.
Recent works have shown that the interplay also produces intriguing superconductor (SC) phases such as
topological crystalline SCs \cite{Ueno13, Zhang13, Fang14, Wang16} and nodal SCs \cite{Yang14, Kobayashi14, Kobayashi16, Micklitz, MicklitzB}.
For example, mirror symmetry enables topological phases in three-dimensional SCs \cite{Ueno13, Zhang13},
and Cu$_x$Bi$_2$Se$_3$ \cite{Sasaki11, Hsieh12L} and UPt$_3$ \cite{Tsutumi13} are known as the candidates.
Nowadays, various topological phases are proposed from systematic topological classification of the quantum matter based on the crystal symmetry
\cite{Chiu13, Morimoto13, Chiu14, Shiozaki14, Shiozaki16, Chiu16}.

Meanwhile, Weyl SCs \cite{Meng12, Sau12, Silaev12, Yang14, Yong14} are three-dimensional SCs with point nodes 
which are stable without protection by crystal symmetries.
The point nodes are called Weyl nodes.
Weyl nodes are protected topologically by monopole charges related to Chern numbers,
and always exist in pairs with opposite monopole charges \cite{Murakami07, Wan11}.
Hence, Weyl nodes cannot appear or vanish unless the pair creation or annihilation occurs.
Because intrinsic particle-hole symmetry gives opposite monopole charges to Weyl nodes at $\bm{k}$ and $-\bm{k}$ \cite{Sato16},
broken time-reversal symmetry is necessary for Weyl SCs
although Weyl semimetals are realizable in time-reversal invariant systems \cite{Murakami07, Halasz12, Ojanen13, Okugawa14, Liu14, Okugawa17}.
The candidates of Weyl SCs are SrPtAs \cite{Biswas13, Fischer14}, uranium-based compounds (including UPt$_3$) \cite{Goswami13, Yamashita15, Goswami15, Yanase16} , 
PrOs$_4$Sb$_{12}$ \cite{Izawa03, Aoki03, Abu07, Kozii16}, Nb$_x$Bi$_2$Se$_3$ \cite{Smylie16, Yuan17, Chirolli17}
and so on.
It is also predicted that an external magnetic field induces a phase transition from a noncentrosymmetric line-node SC phase to a Weyl SC phase\cite{Daido16}.

\begin{figure}[b]
\includegraphics[width=8cm]{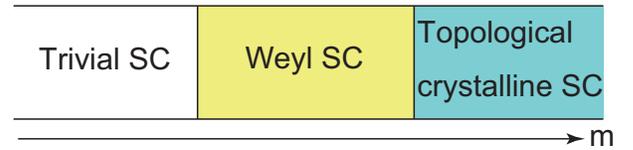}
\caption{\label{gdia}
A generic phase diagram for the Weyl SC and the topological crystalline SC in the absence of time-reversal symmetry.
$m$ is any paramter determining the phase, such as an external magnetic field, a chemical potential, pressure, etc.}
\end{figure}

In this paper, we focus on Weyl superconductivity in mirror-symmetric SCs breaking time-reversal symmetry.
The Weyl SCs are difficult to predict from crystal symmetries as with Weyl semimetals
since their Weyl nodes are located at general points.
Meanwhile, the Weyl semimetal phase necessarily intervenes between trivial and topological insulator phases 
when Kramers degeneracy is absent \cite{Murakami07, Murakami08, Alexandradinata14, Kim16}.
The universal emergence of Weyl semimetal phase has been used to search for candidate materials of the Weyl semimetals \cite{Liu14, Rauch15, Liang17}.
Therefore, it should be useful to study Weyl SC as an intermediate phase in order to search for candidate materials of the Weyl SCs.
In this paper, we show that the Weyl SC phase generally appears between trivial and topological crystalline SC phases in an odd-parity or a noncentrosymmetric SC,
as shown in Fig.~\ref{gdia}.
Also, we investigate how the system enters the topological crystalline SC phase through the Weyl SC phase.
We show that trajectories of the Weyl nodes due to the change in the parameters determine the topological phase transition.

\textit{Weyl SC phase between trivial and topological crystalline SC phases.}
We start by introduction of a topological crystalline SC and a Weyl SC in order to discuss a topological phase transition between the two phases \cite{SUP}.
If a Bogoliubov-de Gennes (BdG) Hamiltonian without time-reversal symmetry has a mirror symmetry,
the system can realize a topological crystalline SC phase \cite{Ueno13, Zhang13, Chiu13, Morimoto13, Shiozaki14}.
When the normal state is mirror-symmetric,
the mirror symmetry is preserved in the BdG Hamiltonian
if the gap function is mirror-odd or even, i.e. $M\Delta (\bm{k}) M^t=\mp \Delta (\bm{k})$ on the mirror plane.
$M$ and $\Delta(\bm{k})$ are mirror operation in the normal state and the gap function of the SC, respectively.
The topological SC is characterized by a mirror Chern number \cite{Teo08}.
The mirror Chern number $\nu ^{(\lambda )}$ is a Chern number defined in the mirror sector  
of the mirror eigenvalue $\lambda =\pm i$ on the mirror plane. 
We denote the BdG Hamiltonian in the mirror sector of $\lambda$ as $\mathcal{H}_{\lambda}(\bm{k})$.
When the SC is gapped and some of mirror Chern numbers are nonzero,
a topological crystalline SC is realized.

To define the mirror Chern number, we introduce Berry connection $\bm{A}^{\lambda}(\bm{k})$ and Berry curvature $\bm{F}^{\lambda}(\bm{k})$
in the mirror sector $\mathcal{H}_{\lambda}(\bm{k})$ given by \cite{Ueno13}
\begin{align}
&\bm{A}^{\lambda}(\bm{k})=i\sum _{n}\bra{u_n^{\lambda}(\bm{k})}\nabla _{\bm{k}}\ket{u_n^{\lambda}(\bm{k})}, \label{BC} \\
&\bm{F}^{\lambda}(\bm{k})=\nabla _{\bm{k}}\times \bm{A}^{\lambda}(\bm{k}),
\end{align}
where $\ket{u_n^{\lambda}(\bm{k})}$ is the $n$-th eigenstate of the BdG Hamiltonian with the mirror eigenvalue $\lambda$.
The sum in $\bm{A}^{\lambda}(\bm{k})$ is taken over the negative energy states.
For example, we take the mirror plane to be the $xy$ plane.
The mirror Chern number is then given by
\begin{align}
\nu ^{(\lambda )}(k_z)=\frac{1}{2\pi}\int dk_xdk_y F_z^{\lambda}, \label{MC}
\end{align}
where $F_z^{\lambda}$ is integrated over the mirror plane $k_z=0$ or $\pi$.
As seen from Eq.~(\ref{MC}), to change the mirror Chern number,
the system needs a gap closing between the negative and the positive energy states with the same mirror eigenvalues.
The properties of the mirror Chern numbers vary according to mirror-parity of the gap function.
In the mirror-odd (mirror-even) SC,
$\nu ^{(+i)}$ and $\nu ^{(-i)}$ are independent (equal)
because each mirror sector has (does not have) its own particle-hole symmetry \cite{Ueno13, Shiozaki14}.

Meanwhile, in Weyl SCs, 
the Weyl nodes are typically created at general points by accidental band touching between two nondegenerate states.
The band touching and emergent nodes can be described by a two-band effective Hamiltonian if all the states in the SC are nondegenerate.
From analysis of the two-band effective Hamiltonian,
the Weyl nodes are allowed only in an odd-parity or a noncentrosymmetric SC\cite{Kobayashi14, Zhao16, Bzduvsek17}.
To study the Weyl nodes in mirror-symmetric SCs,
we also need to consider band touching on the mirror plane.
Then, there are two types of band evolution after the band touching on the mirror plane in the three-dimensional SC,
depending on the mirror eigenvalues of the two states \cite{Murakami17}.
If the two states have the same mirror eigenvalues,
the band touching leads to pair creation of Weyl nodes \cite{SUP}. 
The created Weyl nodes move symmetrically with respect to the mirror plane.
In contrast, if the two states have opposite mirror eigenvalues,
a line node appears on the mirror plane since the two bands do not hybridize \cite{SUP}.

Hereafter, by using the above arguments, let us discuss a topological phase transition to realize topological crystalline SC phases in three-dimensional SCs.
We set any tunable parameter $m$, which governs the topological phase transition in the mirror-symmetric SC without time-reversal symmetry.
Below, we make the following assumptions:
(i) All the bands are nondegenerate in the SC \cite{AM}.
(ii) The mirror symmetries in the SC are invariant by a change of the parameter $m$.
(iii) The SC becomes gapful all over the Brillouin zone within a finite range of the parameter $m$.

\begin{figure}[t]
\includegraphics[width=8cm]{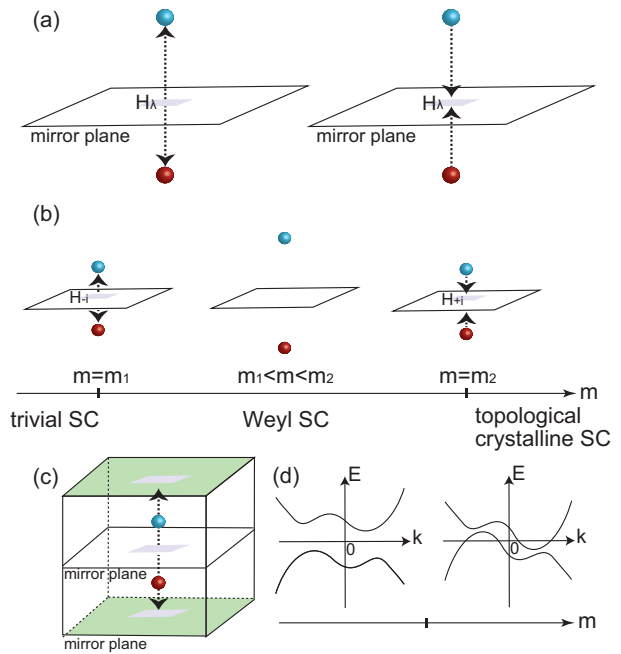}
\caption{\label{pairca}
(a) The pair creation and annihilation of Weyl nodes in the mirror sector $\mathcal{H}_{\lambda}$.
The two colors of the Weyl nodes correspond to opposite monopole charges.
Because of the gap closing on the mirror plane, the mirror Chern number $\nu ^{(\lambda)}$ changes by $\pm 1$.
(b) and (c) Examples of a trajectory formed by one pair of Weyl nodes between trivial and topological crystalline SC phases.
(d) Schematic drawing of the band evolution near zero energy in the absence of inversion and time-reversal symmetries.
The maximum energy of the hole bands can exceed zero energy.
}
\end{figure}

First, we investigate a topological phase transition in odd-parity SCs.
The positive and the negative energy bands at $\bm{k}$ are symmetric with respect to zero energy due to the particle-hole and inversion symmetries.
To see the topological phase transition, we consider the SC gapful and trivial when $m<m_1$.
If the mirror Chern number becomes nonzero while we change the parameter $m$, 
a gap closes between two states with the same mirror eigenvalues on the corresponding mirror plane.
Now, let us assume that the gap closes at $m=m_1$ in the mirror sector $\mathcal{H}_{\lambda}$.
The gap closing not only changes the mirror Chern number $\nu ^{(\lambda )}$ but also leads to pair creation of Weyl nodes [Fig.~\ref{pairca} (a)].
As $m>m_1$ is increased, the Weyl nodes move in the Brillouin zone until the pair annihilation.
Consequently, the SC is in the Weyl SC phase.
Furthermore, we assume that the SC becomes gapped again after a further change of $m$ annihilates all the Weyl nodes at $m=m_2$.
Whether the gapful SC phase in $m>m_2$ is trivial or topological is determined by the trajectory of the Weyl nodes in the Brillouin zone.
To elucidate a relationship between the trajectory and the topological phase transition,
suppose that one pair of Weyl nodes is  created at $m=m_1$ in the $\mathcal{H}_{-i}$ sector on a mirror plane.
Then, the SC system enters the Weyl SC phase with nonzero $\nu ^{(-i)}$.
If the Weyl nodes return to the $\mathcal{H}_{-i}$ sector on the same mirror plane at $m=m_2$,
the mirror Chern number $\nu ^{(-i)}$ becomes zero.
Thus, the SC phase in $m>m_2$ is trivial.
As another example, suppose that the Weyl nodes are pair annihilated at $m=m_2$ in the $\mathcal{H}_{+i}$ sector on the same mirror plane as illustrated in Fig.~\ref{pairca} (b). 
Then, the pair annihilation changes $\nu ^{(+i)}$, whereas $\nu ^{(-i)}$ remains nonzero.
As a result, topological crystalline SC phase is realized.
In this trajectory, the Chern number $\nu = \nu ^{(+i)}+\nu ^{(-i)}$ on the mirror plane is zero in the topological crystalline SC phase.
Hence, mirror symmetry is essential for the topological phase transition.
Moreover, the Weyl nodes can vanish on another mirror plane.
The trajectory also realizes a topological crystalline SC phase with nonzero Chern numbers in addition to nonzero mirror Chern numbers [Fig.~\ref{pairca} (c)].
Generally, there can be more than one pair of Weyl nodes and several mirror planes, and pair creation and annihilation may not occur on the mirror plane.
However, changes of mirror Chern numbers accompany pair creation or annihilation on the mirror plane.
In this way, the generic topological phase transition can be understood from the trajectories formed by all the Weyl nodes.
As a result, the Weyl SC phase can be regarded as an intermediate nodal state between the gapful SC phases with different mirror Chern numbers.

Second, we comment on a topological phase transition between trivial and topological crystalline SC phases in noncentrosymmetric SCs.
Because the BdG Hamiltonian breaks both time-reversal and inversion symmetries,
the maximum energy of the hole bands may be larger than zero \cite{Wong13, Hao16, Daido17}, depending on the parameter $m$ [Fig.~\ref{pairca} (d)].
Hence, Weyl nodes formed by the electron and the hole bands can deviate from zero energy in general.
Then, the mirror Chern numbers in Eq.~(\ref{MC}) are not available for characterization of the topological phase because the Berry connection is defined by negative states.
However, we can use the mirror Chern numbers by replacing the sum of the negative states with that of the hole bands in Eq.~(\ref{BC})
even if the maximum energy of the hole bands exceeds zero energy.
The reason is that the mirror Chern number defined by the hole bands are unchanged as long as the gap survives between the hole and the electron bands.
Therefore, our theory about trajectories of Weyl nodes is also applicable to the topological phase transition in the mirror-symmetric SCs breaking inversion symmetry.

Finally, we show that the mirror-parity of the gap function restricts the trajectories to realize a topological crystalline SC phase.
There is no such restriction on the trajcetory in time-reversal breaking Weyl semimetals
since the mirror-parity is related to particle-hole symmetry \cite{Ueno13, Shiozaki14, Chiu14}.
In the following, we clarify the possible trajectories and the behavior of the gap closing,
which depend on the mirror-parity of the gap function.

We begin with a SC with a mirror-odd gap function.
Since $\nu ^{(+i)}$ and $\nu ^{(-i)}$ are independent, 
the topological crystalline SC phase is realizable from the trajectory as shown in Fig.~\ref{pairca} (b).
Moreover, when inversion symmetry is present,
a gap closing occurs on the mirror plane between the positive and the negative energy states with the same mirror eigenvalues due to particle-hole symmetry in each of the mirror sectors \cite{SUP}.
Namely, the gap closing in the SC necessarily leads to pair creation of Weyl nodes.
Next, we consider a topological phase transition in a SC with a mirror-even gap function. 
Then, $\nu ^{(+i)}$ is always equal to $\nu ^{(-i)}$, unlike the mirror-odd SC.
Thus, pair creation or annihilation in the $\mathcal{H}_{+i}$ sector coincides with that in the $\mathcal{H}_{-i}$ sector.
Hence, if the two pairs of the Weyl nodes emerge from the mirror plane and they return to the same mirror plane,
the gapful SC becomes topologically trivial again.
In order to reach the nontrivial phase, the Weyl nodes need to vanish away from the mirror plane where they have emerged.
Additionally, the gap on the mirror plane can close between the two states with opposite mirror eigenvalues 
since each of the mirror sectors does not keep particle-hole symmetry.
The gap closing then yields a nodal line on the mirror plane.

\textit{Model calculation.}
To demonstrate our theory, we study a SC modeled on a cubic lattice with mirror symmetry.
As an example,
we consider a BdG Hamiltonian written by $H=\frac{1}{2}\sum _{\bm{k}}\Psi ^{\dagger}_{\bm{k}}\mathcal{H}(\bm{k})\Psi _{\bm{k}}$
with $\Psi ^{\dagger}_{\bm{k}}=(c^{\dagger}_{\bm{k}\uparrow}, c^{\dagger}_{\bm{k}\downarrow}, c_{-\bm{k}\uparrow}, c_{-\bm{k}\downarrow} )$ and
\begin{align}
\mathcal{H}(\bm{k})=
\begin{pmatrix}
\xi _{\bm{k}}-Bs_z & \bm{\Delta}(\bm{k}) \\
\bm{\Delta}^{\dagger}(\bm{k}) & -\xi _{\bm{k}}+Bs_z
\end{pmatrix}. \label{modelc}
\end{align}
Here, $\xi _{\bm{k}}=2t_x\cos k_x+2t_y\cos k_y+2t_z\cos k_z-\mu$ is a kinetic energy, 
and $\bm{\Delta}(\bm{k})=i\bm{d}\cdot \bm{s} s_y$ is a gap function,
with $\bm{d}=\Delta (\sin k_x, \sin k_y ,\sin k_z)$ and $\bm{s}=(s_x,s_y,s_z)$ Pauli matrices acting on the spin space.
$B$ is an external magnetic field breaking time-reversal symmetry.
The eigenvalues are
\begin{align}
E(\bm{k})=\pm \Bigl[ \xi _{\bm{k}}^2+B^2+\sum _id_i^2\pm 2B\sqrt{\xi _{\bm{k}}^2+d_z^2}\Bigr] ^{1/2}.
\end{align}
We note that this model describes an odd-parity SC.
This model without the magnetic field is studied as a time-reversal invariant topological SC \cite{Sato09}.

Now, the normal state has a mirror symmetry with respect to the $xy$ plane, and the mirror operation is given by $M_z=-is_z$.
Thus, the gap function is mirror-odd because $M_z\bm{\Delta} (k_x,k_y,k_z)M_z^t=-\bm{\Delta }(k_x,k_y,-k_z)$.
The BdG Hamiltonian also has a mirror symmetry described by $\tilde{M}_z=\mathrm{diag}(M_z, M_z)$.
Therefore, the mirror Chern numbers $\nu ^{(\pm i)}(k_z)$ can be defined on the planes $k_z=0$ and $\pi$ in this model.

On the mirror planes, the Hamiltonian can be block-diagonalized in the diagonal basis of $\tilde{M}_z$.
Each mirror sector of the eigenvalues $\pm i$ is described by
\begin{align}
\mathcal{H}_{\pm i}(\bm{k})=\pm \Delta \sin k_x\tau _x -\Delta \sin k_y \tau _y +[\xi _{\bm{k}}\pm B]\tau _z, \label{mirrorH}
\end{align}
where $\tau _{x,y,z}$ are Pauli matrices.
According to Eq.~(\ref{mirrorH}),
the mirror Chern numbers change when $\xi _{\bm{k}}\pm B=0$ on the mirror planes.

\begin{figure*}[h]
\includegraphics[height=8cm]{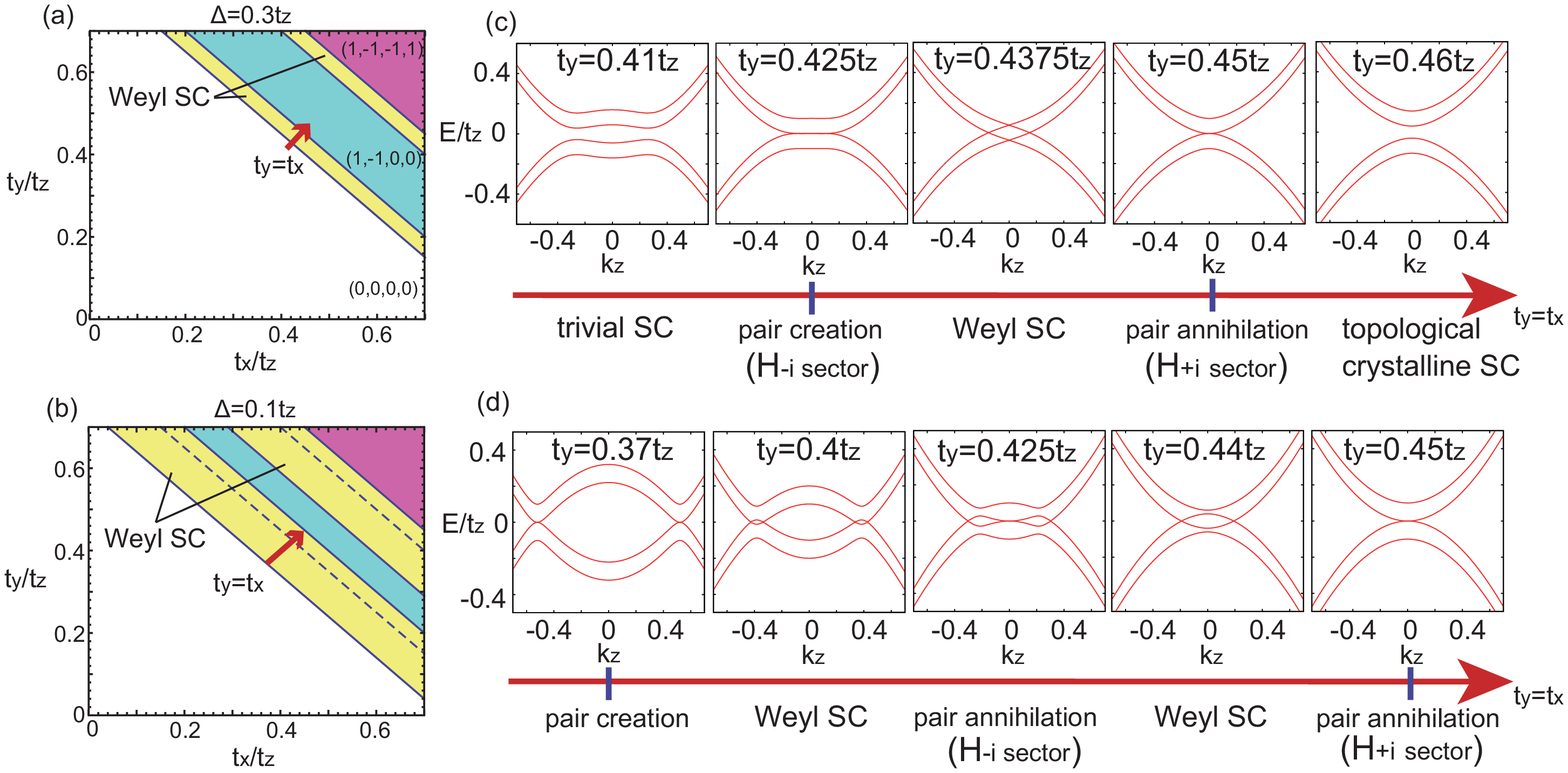}
\caption{\label{pd}
(a) The phase diagram with $\mu =0.25t_z$, $B=0.05t_z$ and $\Delta =0.3t_z$.
The blue and the purple regions are SC phases with the nonzero mirror Chern numbers 
$(\nu ^{(+i)}(0),\nu ^{(-i)}(0),\nu ^{(+i)}(\pi ),\nu ^{(-i)}(\pi ))=(1,-1,0,0)$ and $(1,-1,-1,1)$, respectively.
The yellow regions are Weyl SC phases.
(b) The same as (a) with $\mu =0.25t_z$, $B=0.05t_z$ and $\Delta =0.1t_z$.
The dashed lines represent the parameters where a pair annihilation happens on the mirror plane whereas the system remains in the Weyl SC phase.
(c) and (d) Band evolutions of the SC on the line $\bm{k}=(\pi ,\pi , k_z)$ along the arrows in (a) and (b).
The arrows in the phase diagrams indicate the line $t_y=t_x$.
(c)The mirror Chern number $\nu ^{(-i)}(0) (\nu^{(+i)}(0))$ change at $t_y=0.425t_z (0.45t_z)$. The Weyl node at $\pm k_z (k_z>0)$ has monopole charge $\pm 1$.
(d) The pair creation happens at $t_y=0.37t_z$.
The Weyl nodes at $k_z^1$ and $k_z^2$ $(k_z^1>k_z^2>0)$ have monopole charges $-1$ and $+1$.
}
\end{figure*}

Figure \ref{pd} (a) and (b) show phase diagrams in this model with $\Delta =0.3t_z$ and $0.1t_z$, respectively.
Both of the phase diagrams are obtained when $\mu =0.25t_z$ and $B=0.05t_z$.
For example, we see the topological phase transition along the $t_y=t_x$ line represented by the arrow in the phase diagram of Fig.~\ref{pd} (a). 
The band evolution is shown in Fig.~\ref{pd} (c).
When $t_y=0.425t_z$,
the pair creation happens at $\bm{k}=(\pi, \pi, 0)$ in the $\mathcal{H}_{-i}$ sector.
The Weyl nodes move along the line $\bm{k}=(\pi ,\pi, k_z)$ as $t_y=t_x$ becomes larger.
Eventually, the Weyl nodes are pair-annihilated at $\bm{k}=(\pi, \pi, 0)$ in the $\mathcal{H}_{+i}$ sector when $t_y=0.45t_z$.
The trajectory is identical to that in Fig.~\ref{pairca} (b),
realizing the topological crystalline SC phase.

Moreover, we see a topological phase transition for $\Delta =0.1t_z$.
We also consider band evolution along the $t_y=t_x$ line in Fig.~\ref{pd} (b).
When $t_y=0.37t_z$, four Weyl nodes emerge at points on the $\bm{k}=(\pi ,\pi, k_z)$ line but not on the mirror plane.
When we increase $t_y=t_x$, the one pair vanishes in the $\mathcal{H}_{-i}$ sector, and then the other pair does in the $\mathcal{H}_{+i}$ sector.
This trajectory also realize a topological crystalline SC phase because the two pairs vanish in the different mirror sectors.
Therefore, in both cases, the SC system enters the topological crystalline SC phase via the Weyl SC phase,
which is consistent with our theory.

In the Weyl and the topological crystalline SC phases, Majorana states appear on the surface \cite{Ueno13, Shiozaki14, Chiu14,Meng12, Sau12, Silaev12}.
We show evolutions of the surface states of this model by changing the parameters in the same way as the bulk states in the Supplemental Material \cite{SUP}.

\textit{Conclusion and discussion.}
In the present paper, we have investigated Weyl superconductivity in mirror symmetric superconductors without time-reversal symmetry. 
We have shown that Weyl superconductivity universally emerges between the trivial and the topological crystalline superconductor phases in odd-parity or noncentrosymmetric superconductors.
We have also discussed a relationship between the Weyl nodes and the topological phase transition.
It is shown that trajectories of the Weyl nodes determine the topological phase after the pair annihilation.

Our generic results are applicable to various unconventional superconductors breaking time-reversal symmetry because many crystals have mirror symmetry.
Thus, the theory is useful for prediction of Weyl and topological crystalline superconductors in addition to theoretical construction of the topological phase diagram.
For example, recent papers have implied that an external magnetic field moves Weyl nodes in the Brillouin zone \cite{Yang14, Daido16}.
As shown in our model calculation, changing shapes of Fermi surfaces in the normal state can also induce the topological phase transition, which is expected by doping and pressure.
Hence, the topological phase transition predicted in this paper can be realized by controlling these parameters experimentally. 


\textit{Acknowledgment.}
This work was supported by JSPS KAKENHI Grant No. 16J08552,
Grants-in-Aid for Scientific Research on Innovative Areas ``Topological Materials Science'' (Grant No. JP16H00988)
and ``Nano Spin Conversion Science'' (Grant No. JP17H05179).

\bibliographystyle{apsrev4-1}
\bibliography{WeylSC}

\begin{thebibliography}{73}
\bibitem [{AM()}]{AM}%
  \BibitemOpen
  \href@noop {} {\bibinfo  {journal} {This assumption is realizable by strong
  spin-orbit interaction in noncentrosymmetric systems or by the Zeeman effect,
  for example}\ }\BibitemShut {NoStop}%
\end{thebibliography}
\begin{thebibliography}{19}
\bibitem [{Che()}]{Chernweak}%
  \BibitemOpen
  \href@noop {} {\bibinfo  {journal} {A weak topological invariant $N_{weak}$
  is also introduced \cite{Chiu13}. If $\nu (0)\nu (\pi )>0$,
  $N_{weak}=\mathrm{sgn}(\nu (0))\min (|\nu (0)|,|\nu (\pi )|)$. Meanwhile, if
  $\nu (0)\nu (\pi )<0$, $N_{weak}=0$}\ }\BibitemShut {NoStop}%
\bibitem [{ASS()}]{ASSUMPTION}%
  \BibitemOpen
\bibfield  {journal} {  }\href@noop {} {\bibinfo  {journal} {If the bands are
  doubly degenerate, fourfold-degenerate point nodes with monopole charges can
  emerge in even-parity SCs breaking time-reversal symmetry \cite{Zhao16,
  Kobayashi14, Chiu14}. The point nodes typically have monopole charges $\pm2$
  \cite{Zhao16, Kobayashi14, Chiu14}. However, such cases are beyond our
  scope}\ }\BibitemShut {NoStop}%
\end{thebibliography}
\end{document}



\title{Supplemental material for "Generic phase diagram for Weyl superconductivity in mirror-symmetric superconductors"
}

\author{Ryo Okugawa}
\affiliation{%
 Department of Physics, Tokyo Institute of Technology, 2-12-1 Ookayama, Meguro-ku, Tokyo 152-8551, Japan
}%
\author{Takehito Yokoyama}%
\affiliation{%
 Department of Physics, Tokyo Institute of Technology, 2-12-1 Ookayama, Meguro-ku, Tokyo 152-8551, Japan
}%

\renewcommand{\thefigure}{S\arabic{figure}}
\renewcommand{\theequation}{S\arabic{equation}}
             
\maketitle

\section{Symmetries in superconductors}
To study the topological phase transitions,
we explain particle-hole, mirror, and inversion symmetries in the superconductors (SCs).
The BdG Hamiltonian is generally given by
\begin{align}
&H=\frac{1}{2}\sum _{\bm{k},\alpha ,\beta}
(c^{\dagger}_{\bm{k}\alpha},c_{-\bm{k}\alpha}) 
\mathcal{H}(\bm{k})
\begin{pmatrix}
c_{\bm{k}\beta} \\
c^{\dagger}_{-\bm{k}\beta}
\end{pmatrix}, \\
&\mathcal{H}(\bm{k})=
\begin{pmatrix}
\mathcal{E}_{\alpha \beta}(\bm{k}) & \Delta _{\alpha \beta}(\bm{k}) \\
\Delta ^{\dagger}_{\alpha \beta}(\bm{k}) & -\mathcal{E}^t_{\alpha \beta}(-\bm{k})
\end{pmatrix},
\end{align}
where $\mathcal{E}(\bm{k})$ is the normal-state Hamiltonian measured from the chemical potential,
and $\Delta (\bm{k})$ is the gap function.
The subscripts $\alpha$ and $\beta$ indicate spin and orbital indices.
In general, the normal-state Hamiltonian and the gap function are $N\times N$ matrices.
Because of the fermion anticommutation relations,
the gap function satisfies
$\Delta _{\alpha \beta}(\bm{k})=-\Delta _{\beta \alpha}(-\bm{k})$.
Thus, the BdG Hamiltonian has particle-hole symmetry ${C}$:
\begin{align} 
&{C}\mathcal{H}(\bm{k}){C}^{-1}=-\mathcal{H}(-\bm{k}), \\
&{C}=
\begin{pmatrix}
0 & \bm{1} \\
\bm{1} & 0
\end{pmatrix}K,
\end{align}
where $\bm{1}$ is an $N\times N$ unit matrix, and $K$ is a complex conjugate operator.
Particle-hole symmetry relates the energy eigenvalues at $\bm{k}$ and $-\bm{k}$.

Next, we give a relationship between the BdG Hamiltonian and crystal symmetries \cite{Ueno13, Kobayashi14, Shiozaki14}.
When the normal state has a crystal symmetry $g$ which transforms $\bm{k}$ as $\bm{k}\rightarrow D(g)\bm{k}$ with a unitary matrix $D(g)$,
the BdG Hamiltonian can also have the symmetry depending on the gap function.
Let $g$ be inversion $(P)$ or mirror $(M)$ symmetry of the normal state.
By assumption, the normal-state Hamiltonian and $g$ are related by
\begin{align}
U_g\mathcal{E} (\bm{k})U_g^{-1}=\mathcal{E} (D(g)\bm{k}),
\end{align}
where $U_g$ is a matrix representation of $g$ in the basis of the normal-state Hamiltonian.
If the gap function is transformed by $g$ as
\begin{align}
&U_g\Delta (\bm{k})U_g^t=\eta _g\Delta (D(g)\bm{k}), \\
&\eta _g=\pm 1,
\end{align}
the BdG Hamiltonian preserves the symmetry $g$.
Namely, the BdG Hamiltonian satisfies
\begin{align}
&\tilde{U}_g^{\eta _g}\mathcal{H}(\bm{k})[\tilde{U}_g^{\eta _g}]^{-1}=\mathcal{H}(D(g)\bm{k}), \\
&\tilde{U}_g^{\eta _g}=
\begin{pmatrix}
U_g & 0 \\
0 & \eta _gU_g^{\ast}
\end{pmatrix}.
\end{align}
Particularly, when $\eta _{g}=1 (-1)$, particle-hole symmetry commutes (anticommutes) with $g$,
which are expressed by
\begin{align}
{C}\tilde{U}_g^{\eta _g}=\eta _g\tilde{U}_g^{\eta _g}{C}. \label{CU}
\end{align}

When $g=P$ and $\eta _P=1 (-1)$,
the SC is named an even-parity (odd-parity) SC.
In the inversion-symmetric SCs, we obtain $({C}\tilde{U}_P^{\eta _P})^2=\eta _P$ from Eq.~(\ref{CU}).
$U_P=1$ and $\eta _P=-1$ are used for the model in the main text.
Meanwhile, when $g=M$ and $\eta _M=1 (-1)$,
we call the SC mirror-even (mirror-odd) SC \cite{Ueno13, Kobayashi14, Shiozaki14}.
In the main text, $M$ and $U_M$ are not distinguished for brevity.
Hereafter, the matrix representation $U_g$ is similarly written as $g$.
We also denote $\tilde{U}_g^{\eta _g}$ as $\tilde{g}$,
or as $\tilde{g}^{\eta}$ with $\eta =\pm$ if necessary. 

If the BdG Hamiltonian is mirror-symmetric, 
mirror Chern numbers $\nu ^{(\pm i)}(k_z)$ can be defined \cite{Teo08, Ueno13}.
Moreover, 
a strong topological invariant $N_{M\mathbb{Z}}$ can be introduced in the three-dimensional gapped system.
$N_{M\mathbb{Z}}$ is given by \cite{Chiu13, Chiu14}
\begin{align}
N_{M\mathbb{Z}}=\mathrm{sgn}(\nu (0)-\nu (\pi ))(|\nu (0)|-|\nu (\pi )|),
\end{align}
where $\nu (k_z)$ is one of $\nu ^{(\pm i)}(k_z)$ \cite{Chernweak}.
It is seen that changes of $N_{M\mathbb{Z}}$ follow those of the mirror Chern numbers $\nu ^{(\pm i)}$.
Therefore, we focus on the changes of the mirror Chern numbers $\nu ^{(\pm i)}$ in this paper
because they are directly related to pair creation and annihilation of Weyl nodes.

\section{Gap closing in mirror-symmetric superconductors without time-reversal symmetry}
\subsection{Gap closing in noncentrosymmetric SCs}\label{const}
We construct an effective Hamiltonian to study the topological phase transition involving the Weyl SC phase.
We assume that all the bands are nondegenerate \cite{ASSUMPTION}, and that time-reversal symmetry is broken.
Thus, we can describe a gap closing at $\bm{k}=\bm{k}_0$ by a two-band effective Hamiltonian.
The behavior of the gap closing can be determined by symmetries which leave $\bm{k}=\bm{k}_0$ invariant \cite{Murakami17}.
In noncentrosymmetric SCs, the two-band effective Hamiltonian can be written by
\begin{align}
H_{eff}(\bm{q},m) =a_0(\bm{q},m)\sigma _0 +a_1(\bm{q},m)\sigma _1+a_2(\bm{q},m)\sigma _2 +a_3(\bm{q},m)\sigma _3, \label{effH}
\end{align}
where $\bm{q}=\bm{k}-\bm{k}_0$, and $m$ is a tunable parameter.
Here, $\sigma _{0,1,2,3}$ are Pauli matrices, 
and $\sigma _3=+1$ and $-1$ indicate the band indices.
To see the gap closing, we assume that the gap is open within $m<m_0$,
and that the gap closes at $\bm{q}=0$ and $m=m_0$.
The band evolution within $m>m_0$ is determined by the number of the parameters necessary for the gap closing \cite{Murakami17}.
In three-dimensional systems, a topological phase with stable point (line, surface) nodes generally emerges when $m>m_0$,
provided that difference between the number of variables and conditions for the gap closing is 1 (2, 3).
For example, we consider a gap closing at general points.
Then, there is no constraint by symmetries on the effective Hamiltonian.
Because we have four variables $(\bm{q} ,m)$ in the three-dimensional system,
three conditions ($a_{1,2,3}=0$) should be satisfied. 
Thus, if the gap closing occurs at $(\bm{q}, m)=(0,m_0)$,
a pair of Weyl nodes is created \cite{Murakami07, Murakami08, Murakami17}.

Now because we consider the topological phase transition in mirror symmetric SCs,
we need to clarify a behavior of a gap closing on the mirror plane.
For simplicity, we assume that the mirror plane is the $xy$ plane.
Then, the effective Hamiltonian satisfies
\begin{align}
\tilde{M}H_{eff}(q_x, q_y, -q_z, m)\tilde{M}^{-1}=H_{eff}(q_x, q_y, q_z, m). \label{effm}
\end{align}

The structures of the nodes after the gap closing on the mirror plane vary depending on mirror eigenvalues of the two states.
If the two states have the same mirror eigenvalues, i.e., $\tilde{M}=\pm i\sigma _0$,
Eq.~(\ref{effm}) becomes $H_{eff}(q_x, q_y, -q_z, m)=H_{eff}(q_x, q_y, q_z, m)$.
Because this equation is satisfied automatically at $q_z=0$,
this condition does not affect the gap closing on the mirror plane $(q_z=0)$.
Therefore, after the gap closes between the two states with the same mirror eigenvalues at $m=m_0$,
Weyl nodes appear symmetrically with respect to the mirror plane \cite{Murakami17}.

On the other hand, if the two states have the opposite mirror eigenvalues, i.e., $\tilde{M}=\pm i\sigma _3$,
Eq.~(\ref{effm}) is reduced to $\sigma _3H_{eff}(q_x, q_y, -q_z, m)\sigma _3=H_{eff}(q_x, q_y, q_z, m)$.
As a result, we obtain $H_{eff}(q_x, q_y, 0, m)=a_0(q_x,q_y,0,m)\sigma _0+a_3(q_x, q_y ,0, m)\sigma _3$ on the mirror plane ($q_z=0$).
In this case, the mirror symmetry imposes only one condition ($a_3=0$),
while we have three variables $(q_x, q_y ,m)$ on the mirror plane.
Hence, a line node appears on the mirror plane after the gap closing \cite{Murakami17}.

\subsection{Gap closing in inversion-symmetric SCs}
If SCs have inversion symmetry, positive and negative energy states at $\bm{q}$ are related by combination of particle-hole and inversion symmetries.
The effective Hamiltonian also satisfies
\begin{align}
C\tilde{P}H_{eff}(\bm{q})(C\tilde{P})^{-1}=-H_{eff}(\bm{q}).
\end{align}
Therefore, we need to take $C\tilde{P}$ symmetry into account when we construct the effective Hamiltonian of the inversion-symmetric SCs.

Firstly, we consider odd-parity SCs.
Since $(C\tilde{P}^{-})^2=-1$, $C\tilde{P}^{-}=-i\sigma _2K$ can be chosen \cite{Zhao16},
which leads to $a_0(\bm{q})=0$.
Then, the effective Hamiltonian becomes
\begin{align}
H_{eff}(\bm{q},m) =a_1(\bm{q},m)\sigma _1+a_2(\bm{q},m)\sigma _2 +a_3(\bm{q},m)\sigma _3. \label{effCP1}
\end{align}
From Eq.~(\ref{effCP1}), the odd-parity SCs have the same conditions for the gap closing as the noncentrosymmetric SCs.
Hence, we can apply the same discussion on the gap closing as used in Sec. \ref{const}.
We note that the gap closes and the Weyl nodes appear at zero energy by $C\tilde{P}$ symmetry.
In addition, we show that $C\tilde{P}$ symmetry determines mirror operator $\tilde{M}$ which acts on the effective Hamiltonian.
In mirror-symmetric SCs, the mirror symmetry satisfies $\tilde{M}^{\eta}C=\eta C\tilde{M}^{\eta}$.
We assume that the positive-energy state $\ket{E_+, \lambda}$ has a mirror eigenvalue $\lambda =\pm i$ on the mirror plane.
Then, the negative-energy state $C\tilde{P}\ket{E_+, \lambda}$ has a mirror eigenvalue $-\eta \lambda$ because
\begin{align}
\tilde{M}^{\eta}C\tilde{P}\ket{E_+, \lambda}=\eta C\tilde{P}\tilde{M}^{\eta}\ket{E_+, \lambda}=-\eta \lambda C\tilde{P}\ket{E_+, \lambda}.
\end{align}
Thus, the mirror operation is represented as $\tilde{M}^{\eta}=\lambda ~\mathrm{diag}(1, -\eta)$.
In other words, when the gap function is mirror-odd (even),
the mirror eigenvalues of the two states are the same (different).
Hence, the gap closing on the mirror plane leads to pair creation of Weyl nodes in the odd-parity SCs with the mirror-odd gap function.
This is consistent with the fact that each mirror sector has its own particle-hole symmetry in the mirror-odd SC.
On the other hand, in the odd-parity SCs with the mirror-even gap function,
a nodal line appears from the gap closing on the mirror plane.

Secondly, we construct the effective Hamiltonian for even-parity SCs.
Then, $C\tilde{P}^+=\sigma _1K$, which yields $(C\tilde{P}^+)^2=1$, can be chosen \cite{Zhao16}.
Thus, because we obtain $H_{eff}(\bm{q},m)=a_3(\bm{q},m)\sigma _3$ from Eq.~(\ref{effH}),
there is one condition for the gap closing.
As a result, the gap closing produces a surface node \cite{Kobayashi14, Agterberg17, Timm17, Bzduvsek17, Turker17}. 
This case is not treated in the main text 
because we are interested in the Weyl SCs.

\section{Evolution of Majorana surface states: from Majorana arc to Majorana cone}
Topological SCs generally exhibit surface states protected topologically.
Therefore, we here discuss evolutions of the topological surface states and the bulk states
in a topological phase transition between Weyl and topological crystalline SC phases.
To understand how the surface states evolve, we use the SC model with the mirror-odd gap function described by Eq.~(4) in the main text.
In general, Weyl SCs have Majorana arc states on the surface \cite{Meng12, Sau12, Silaev12}.
The Majorana arc states lie between projections of Weyl nodes with opposite monopole charges onto the surface. 
On the other hand, topological crystalline SCs show Majorana surface states
when the gap function is mirror-odd \cite{Ueno13, Shiozaki14, Chiu14}.            

\begin{figure*}[h]
\includegraphics[width=17cm, height=12cm]{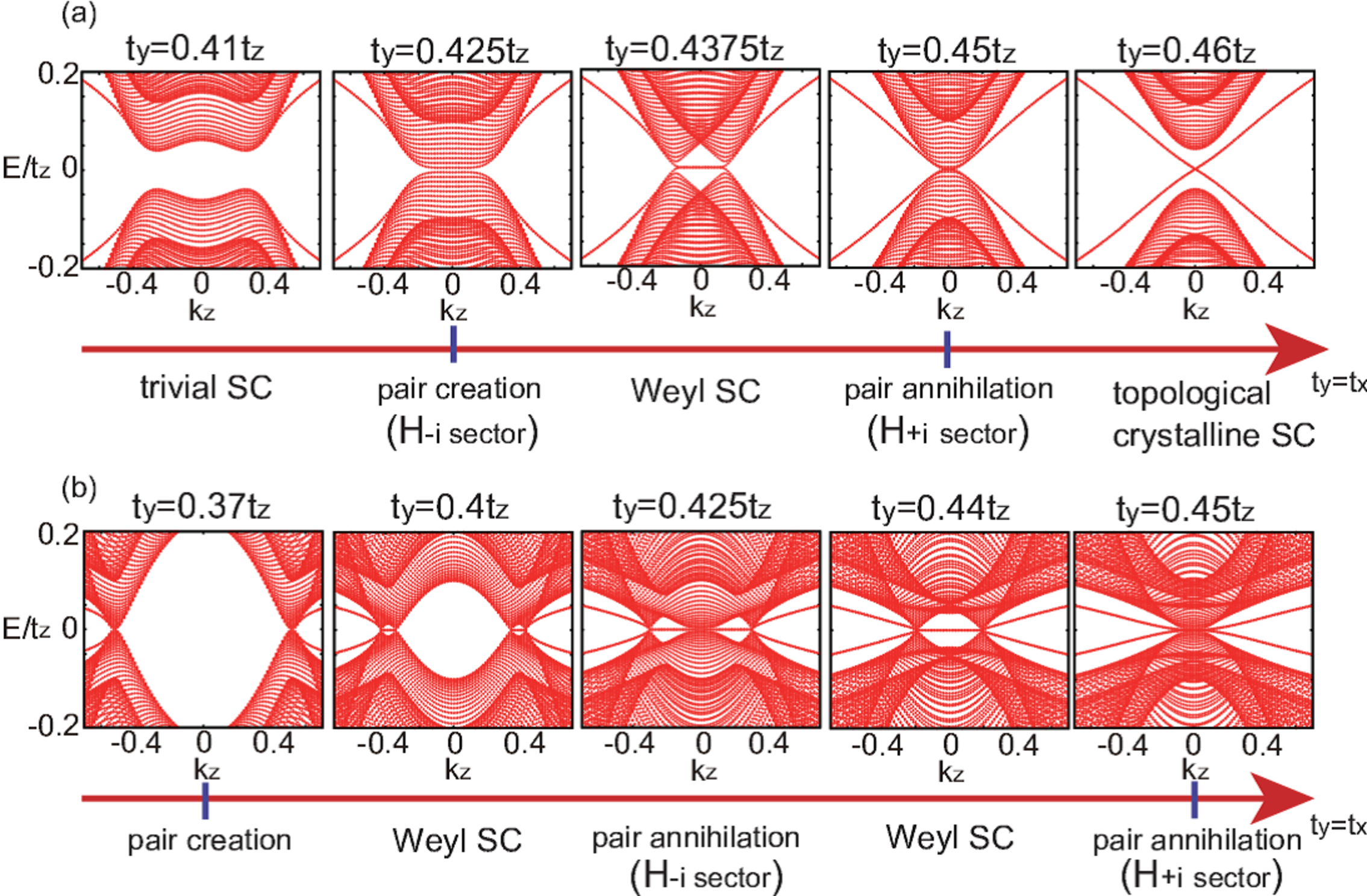}
\caption{\label{sev}
(a) and (b) Surface band evolutions on the line $\bm{k}=(\pi ,k_z)$ along the arrow $t_y=t_x$ in Fig.~ 3 (a) and (b) in the main text.
}
\end{figure*}

We calculate surface states for a slab with a (100) surface to preserve mirror symmetry of the model.
We choose the same parameters $\mu =0.25t_z$ and $B=0.05t_z$ as used in the main text [Fig.~3 (a) and (b)].
We also investigate the evolution of the surface states along $t_y=t_x$ line described by the arrows in the phase diagrams.
Figure \ref{sev} (a) and (b) show evolutions of the surface bands on the line $\bm{k}=(k_y=\pi, k_z)$ of the model when $\Delta =0.3t_z$ and $0.1t_z$, respectively.
First, we see the band evolutions of the surface states when $\Delta =0.3t_z$, as shown in Fig.~\ref{sev} (a).
If the system enters into the Weyl SC phase from the trivial SC phase, 
Majorana arc states appear between the projections of the Weyl nodes which have emerged from the mirror plane.
The Majorana arc extends after the phase transition between the trivial and the Weyl SC phases.
As the system approaches the topological crystalline SC phase,
the Majorana arc gradually shortens.
When the phase transition occurs between the Weyl and the topological crystalline SC phases,
the Majorana arc disappears by pair annihilation of the Weyl nodes.
Nevertheless, the Majorana surface states survive because the mirror Chern numbers are nonzero on the bulk $k_z=0$ plane.

Next, we consider the evolution of the surface states when $\Delta =0.1t_z$ [Fig.~\ref{sev} (b)].
Because the four Weyl nodes emerge in the bulk in the Weyl SC phase, two Majorana arcs can be found for each pair.
When we increase the parameter along $t_y=t_x$ line, 
the four Weyl nodes move to the mirror plane $k_z=0$.
The four Weyl nodes vanish on the mirror plane when the SC shows a transition from the Weyl SC phase to the topological crystalline SC phase.
Then, the emergent Majorana arcs disappear by the pair annihilations on the mirror plane.
Eventually, the Majorana surface states remain at $k_z=0$ in the topological crytalline SC phase.
Therefore, we can see that the shrinking Majorana arcs turn into the Majorana states in the topological crystalline SC phases, 
although the trajectories of the Weyl nodes are different for $\Delta=0.3t_z$ and $0.1t_z$.

\bibliographystyle{apsrev4-1}
\bibliography{WeylSC}